\documentclass[10pt,conference,twocolumn]{article}

\usepackage{times,amsmath}
\usepackage{graphicx}
\usepackage{caption}
\usepackage{subcaption}

\usepackage{enumerate}
\usepackage{multirow}
\usepackage{authblk}

\title{Hash in a Flash: Hash Tables for Solid State Devices}

\author[1]{Tyler Clemons}
\author[1]{S M Faisal}
\author[2]{Shirish Tatikonda}
\author[3]{Charu Aggarawl}
\author[1]{Srinivasan Parthasarathy}

\affil[1]{Department of Computer Science,The Ohio State University}
\affil[ ]{\textit {\{clemonst,faisal,srini\}@cse.ohio-state.edu}}

\affil[2]{IBM Almaden Research Center, San Jose, CA, USA}
\affil[ ]{\textit {\{statiko\}@us.ibm.com}}

\affil[3]{IBM T. J. Watson Center, Yorktown Heights, NY, USA}
\affil[ ]{\textit {\{charu\}@us.ibm.com}}
\date{}
\begin{document}
\maketitle
\begin{abstract} 
In recent years, information retrieval algorithms have taken center stage for extracting 
important data in ever larger datasets.  During computation, these datasets are both stored in 
memory and on disk.  As datasets become larger demand for superior storage solutions will rise.  
Advances in hardware technology have lead to 
the increasingly wide spread use of flash storage devices. Such devices have 
clear benefits over traditional hard drives in terms of latency of access,
bandwidth and random access capabilities particularly when reading data.  Thus traditional informational retrieval algorithms, such as TF-IDF, can greatly benefit.
There are however some interesting trade-offs to consider when leveraging the
advanced features of such devices.
On a relative scale 
writing to such devices can be expensive.
This is because typical flash devices (NAND technology) are
updated in blocks. A minor update to a given block requires the
entire block to be erased, followed by a re-writing of the block. On
the other hand, sequential writes can be two orders of magnitude
faster than random writes.  In addition, random writes are degrading
to the life of the flash drive, since each block can support only a
limited number of erasures. 
TF-IDF can be implemented using a counting hash table.  In general, hash tables are a particularly
challenging case for the flash drive because  this data structure
is inherently dependent upon the {\em randomness} of the hash
function, as opposed to the spatial locality of the data. This makes
it difficult to avoid the random writes incurred during the construction of the counting hash table for TF-IDF.  
In this paper, we will study the design landscape for the development of a hash table for flash storage
devices. We demonstrate how to effectively design a  hash table with two related
hash functions, one of which exhibits a data placement property with
respect to the other. 
Specifically, we focus on three designs based on this general philosophy and evaluate
the trade-offs among them along the axes of query performance, insert and update times and
I/O time through an implementation of the TF-IDF algorithm.
\end{abstract}

\section{Introduction}
\label{sec:Intro}
In recent years, advances in hardware technology have led to the
development of flash devices.  These devices  have several
advantages such as  faster seek times because of   a lack of moving
parts. Furthermore, they are more energy-efficient because of their
use of non-mechanical techniques for data storage
\cite{agrawal2,birrell}. Such drives are extremely fast for random
read operations since they do not require the mechanical seeks
necessary for disks.

While data access is extremely fast, the writes to the drive can
vary in speed depending upon the scenario.  Sequential writes are
quite fast, though random writes can be over two orders of magnitude
slower.  The reason for this is the low level of granularity of
erasing and updating data on disks. In fact, depending on the
management of the Flash Transfer Layer (FTL), random updates  in a
flash drive may be significantly slower than random writes on disk
in spite of the fact that the flash drive does not have any moving
parts. On the other hand, random reads are extremely fast on the
flash drive and are almost as fast as sequential reads.  A second
property of the flash drive is that it supports only a  finite
number of erase-write cycles, after which the blocks on the drive
may wear out.

The different trade-offs in read-write speed leads to a number of
challenges for database applications, especially those in which
there are frequent updates to the underlying data. As a result,
there has been a considerable amount of research
\cite{agrawal,agrawal2,birrell,kang,lee1,lee,li} on database
operations in flash storage devices. In particular, index structures
are a challenge for the case of the flash drive because of their
frequent updates with individual records. Such updates can be
expensive, unless they can be carefully batched with the use of
specialized update techniques.  The idea is to minimize the block
overhead in random writes. This approach also reduces the number
of erase-write cycles on the flash, which increases its effective
life.

The hash table is a  widely used data structure in modern database
systems \cite{cormen}. A hash table relies on a hash function to map
{\it keys} to their associated {\it values}. In a well designed
table, the cost of insertion and lookup requires constant
(amortized) time, and is independent of the number of entries in the
table. Such hash tables are commonly used for lookup, duplicate
detection,  searching and indexing in a wide range of domains
including database indexing. A {\it counting hash table} is one in
which in addition to the value associated with a key, a (reference)
count  is also kept up to date in order to keep track of the
occurrences of a specific key-value pair. Counting hash tables are
also widely used. In the programming languages and software
engineering context, such tables can be used for object reference
counting to aid in garbage collection and memory leak detection
activities (e.g. Java JVM \cite{friedman,levanoni}). In the data
mining context, such tables are often used to efficiently count the
number of occurrences of a given pattern (e.g. frequent pattern
mining\cite{relue}).  In the database context, such tables are used for
indexing (e.g. XML indexing, and selectivity
estimation\cite{aboulnaga}).  In the computational linguistics and
information retrieval context, such tables can be used to
efficiently count the number of distinct words and the number of
occurrences per word within a corpus or document collection
\cite{zweig}.

TF-IDF \cite{tfidf3}, Term Frequency{\em -}Inverse Document Frequency, is a common technique
used in text mining and information retrieval \cite{tfidf2}.  TF-IDF measures the importance of a particular word to a document given a corpus of documents.  Words that appear frequently, often referred to as {\em stop-words} e.g. $the$, are given a lower TF-IDF score than words that are more rare as they are assumed to offer more information about a document's subject e.g. $Macintosh$.  For query processing, such as a search engine, documents can be ranked by their relevancy using TF-IDF;  the relevancy of a document increases if a word contained in a query has a high TF-IDF score.  TF-IDF can also be used for document similarity.  A set of keywords can be defined for each document; keywords are defined by those words with a TF-IDF score higher than a set threshold.  Using a similarity measure between the resulting TF-IDF vectors of two documents can yield a similarity score between two documents.  Computing the TF-IDF scores requires accumulating the occurrences of a term; this is an excellent application for counting hash tables.

Hash tables are an enormous challenge for the flash drive because
they are naturally based on random hash functions and exhibit poor
access locality.  Thus, the key property of the hash table,
randomness, becomes a liability on the flash drive. This paper will
provide an effective method for updates to the tables in flash
memory, by using a carefully designed scheme which uses  two closely
related hash functions in order to ensure locality of the update
operations. Specifically, we will be designing a counting hash
table.  Counting hash tables pose an additional challenge since,
unlike standard hash tables, a duplicate key-value pair requires an
in-place update to the specific location. In-place updates are
non-trivial and given their unpredictable nature, they place an
additional burden beyond just insertions and lookups.

This paper is organized as follows. The remainder of this section
will discuss the properties of the flash table which are relevant to
the effective design of the hash table. We will then discuss related
work and the contributions of this paper. In Section \ref{sec:overview}, we will
discuss our different techniques.  Section \ref{sec:experiments} contains the experimental results.
The conclusions and summary are contained in Section \ref{sec:conclusions}.

\subsection{Properties of the Flash Drive}
The solid state drive (SSD) is
implemented with the use of Flash memory; which comes in two types:
NOR and NAND chips. NOR chips are faster and have higher lifetime but NAND chips have higher capacity and have been adopted in most commodity mobile devices using SSDs.

The most basic unit of access is a page which contains between 512 and 4096 bytes,
depending upon the manufacturer. Furthermore, pages are organized
into {\em blocks} each of which may contain  between 32 and 128
pages.  The data is read and written at the level of a page, with the additional
constraint that when any portion of data is
overwritten at a given block, the entire block must be copied to
memory, erased on the flash,  and then copied back to the flash
after modification.  This process is performed automatically by the
software known as the Flash Translation Layer (FTL) on the flash
drive. Thus, even a small random update of a single byte could lead
to an erase-write of the entire block. Similarly, an erase, or clean, can only be performed at the
block level rather than the byte level. Since random writes will
eventually require erases once the flash device is full, it implies
that such writes will require block level operations.  Thus, the
overhead for the case of random writes can be very large, unless one
is careful about the techniques used for data modification. On the
other hand, sequential writes on the flash are quite fast; typically
sequential writes are two orders of magnitude faster than random
writes.

Another technological limitation of the flash drive is that it can
typically support only a limited number of erase-writes. After this,
the blocks on the flash may degrade and they may not be able to
support further writes. Typically, a flash drive can support between
10,000 to 100,000 erase writes. In this respect, random writes are
extremely degrading to the flash because they may trigger many erase writes.
Therefore, it is essential to batch as many updates as possible on blocks.
This is particularly difficult for the case of the hash table because it often contains a
large fraction of cells which are not updated and the writes are
inherently random in nature.

\subsection{Related Work}
Flash devices  have recently found increasing interest in the
database community because of their fast random read and sequential
write performance. Flash devices have been used in enterprise
database applications \cite{lee}, as a write cache to improve
latency \cite{graefe}, and it is also used as an intermediate
structure to improve the efficiency of migrate operations in the
storage layer \cite{koltsidas}. Methods for page-differential
logging for efficiently storing data on flash devices in a DBMS
independent way are discussed in \cite{KimWS10}. Other database
applications such as the design of dynamic self-tuning databases and
the maintenance of database samples have been discussed in
\cite{nath,nath2}.  There has also been work on  designing tree-indexes on  raw flash devices
\cite{kang,cwu} and indexes to deal with the random-write problem \cite{li}.

Rosenblum and Ousterhout proposed the notion of log-structured disk storage
management~\cite{rosenblum1992design} that relies on the assumption that the reads are
cheap (as they are served from memory) and the writes are expensive (due to disk seeks and
rotations).  Not surprisingly, mechanisms similar to log-structured file systems are adopted
in modern SSDs either at the level of FTL or at the level of file system to handle issues
related to wear-leveling and erase-before-write~\cite{dai2004elf,hyun2010,kim2002space,
kim2009flashlite,woodhouse2001jffs}. As we discuss later, some of our buffering strategies are also inspired from log-structured file systems. Our design
exploits the strength of flash-based storage devices in fast sequential writes, and tries
to alleviate the problem of random writes.

There have been hash tables designed with SSDs including the work presented in \cite{cams} in the context of data intensive networked systems, \cite{fawn} in the context of wimpy nodes, \cite{cstash} is in the context of data de-duplication, \cite{mhash} energy efficient memory sensors, and \cite{fstore} persistent storage as write and/or read caches.  In \cite{faisal1} designs a tree index for flash devices.  However, \cite{faisal1} does not address duplicate keys thus it cannot handle a counting hash.

In this work, we design a {\em counting} hash table that maintain frequencies, and this has not been addressed thus far. We store the hash table on the SSD which is not seen in the designs of~\cite{fawn,fstore}. Unlike most of the existing strategies that rely on simple memory-based buffering schemes, we design a novel combination of memory- and disk- based buffering scheme. Our method leverages the strengths of SSDs (fast sequential/random reads, fast sequential writes) to  effectively address the weaknesses in SSDs (random writes, write endurance).  We would like to emphasis that the works presented in this section {\em do not handle a counting hash table}, which is required by algorithms like TF-IDF, but our proposed hash table designs will.
\subsection{Contributions of this paper}
In this paper, we will design a hash table for the SSD. The
hash table is a particularly challenging case compared to the case
of  conventional index structures because it is inherently
dependent upon randomness as opposed to spatial locality. Index
structures, which are dependent upon spatial locality, are much easier
to update because the spatial locality can be leveraged to perform
block updates of particular regions of the index. This is
non-trivial for the case of the hash table in which the  randomness
guarantees that successive updates may occur at completely
unrelated places on the hash table. As a result, it is much more
difficult to cluster updates for the purpose of block updates in a
hash table because successive updates may occur at widely unrelated
places on the hash table.
In this work, we make the following specific contributions --
$(i)$ We propose a mechanism to support large {\em counting} hash tables on SSDs via a two-level hash function, which ensures that the random update property of flash devices is effectively handled, by localizing the updates to SSD;
$(ii)$ We devise a novel combination of memory- and disk- based buffering scheme that effectively addresses the problems posed by SSDs (random writes, write endurance). While the memory-resident buffer leverages the fast random accesses to RAM; the disk-resident buffer exploits fast read performance and fast sequential/semi-random write performance of SSDs;
$(iii)$ We perform a detailed empirical evaluation to illustrate the effectiveness of our approach by demonstrating the TF-IDF algorithm using our hash table.

\section{A Flash-Friendly Hash Table}
\label{sec:overview}

In this section, we will introduce a hash table which is optimized
for flash storage devices. We will introduce a number of different
schemes for implementing the hash-table as well as basic hash table operations on these designs.

The major property of a hash table is that its effectiveness is
highly dependent upon updates which are distributed randomly across
the table.   On the other hand, in the context of a flash-device, it
is precisely this randomness which causes random access to different
blocks of the SSD. Furthermore, updates which are distributed
randomly over the hash table are extremely degrading in terms of the
wear properties of the underlying disk. This makes hashing
particularly challenging for the case of flash devices.

 Hash table addressing is of
two types: {\em open} and {\em closed}, depending upon how the data
is organized and collisions are resolved.  These two kinds of tables
are as follows.
\begin{itemize}
\item {\bf Open Hash Table:} In  an open hash table, each slot of the hash table corresponds
to multiple data entries.  Thus, each slot is a container of entries
which map onto that value of the hash function. Each entry of the
collection is a key and frequency pair.

\item {\bf Closed Hash Table:} In a closed hash table, the entries
are accommodated within the hash table itself. Thus, the hash table
slot contains the hashed string and its frequency.  However, since
multiple objects cannot be mapped onto the same entry, we need a
collision resolution process, when a hashed object  maps onto an
entry which has already been filled. In such a case, a common
strategy  is to use {\em linear probing} in which we cycle through
successive entries of the hash table until we  either find an
instance of the object itself (and increase its frequency), or we
find an empty slot in which we insert the new entry.  We note that a
fraction of the hash table (typically at least a quarter)  needs to
to be empty in order to ensure that the probing process is not a
bottleneck. The fraction of hash table which is full is denoted by
the load factor $f$. It can be shown that $1/(1-f)$ entries of the
hash table are accessed on the average in order to access or update
an entry.
\end{itemize}

In this paper, we will use a combination of the open and closed hash
tables in order to design our update structure. We will use a closed
hash table as the {\em primary hash table} which is stored on the
(Solid State) drive, along with an {\em update hash table} which is
open and available in main memory. The hash functions of the two
tables are different (since the number of entries in the secondary
hash table are much lower than the first), but they are related to
one another in a careful way, so as to guarantee locality of
updates.  We will discuss this slightly later. The secondary hash
table is updated for each incoming record; from time to time,
portions of the secondary table are used in order to update the
primary table in batch mode. The batch-updates are scheduled in a
way so as to minimize the wasteful erase-writes in the update
process.

We assume that the primary hash table contains $q$ entries, where
$q$ is dictated by the maximum capacity planned for the hash table
for the application at hand.   The secondary hash table contains
$\lceil q/r \rceil$ entries where $r<< q$.  The hash function for the primary and
secondary hash tables are denoted by $g(x)$ and $s(x)$, and are
defined  as follows:
\begin{equation}
g(x)= (a\cdot x +b) \mbox{mod} (q)\\
\end{equation}
\begin{equation}
s(x) = ((a \cdot x +b) \mbox{mod} (q)) \mbox{div} (r)
\end{equation}
In general, the scheme will work with any pair of hash functions
$g(x)$ and $s(x)$ which satisfy the following relationship:
\begin{equation}
s(x) = g(x) \mbox{div} (r)
\end{equation}
It is easy to see that the entries which are pointed to by a single
slot of the memory-resident table  are located approximately
contiguously on the drive-resident (closed) table, because of the
way in which the linear probing process works. This is an important
observation, and will be used at several places in ensuring the
efficiency of the approach.   Linear probing essentially assumes that items that collide
onto the same hash function value  will be {\em contiguously located
in a hash table with no empty slots between them}.   Specifically, the $m$th slot on the
secondary table, corresponds to entries starting from $r \cdot (m-1)
+1$ upto entry $ r \cdot m$ in the primary table. We note that {\em
most} entries which would be pointed to by the $m$th slot of the
secondary table would also map onto the afore-mentioned entries in
the primary table, though this would not always be true because of
the overflow behavior of  the linear probing process beyond these
boundaries.

\subsection{Desirable Update Properties of an SSD-based Hash-Table}
In this subsection, we will provide a broad overview of some of the
desirable update properties of a hash-table. In later subsections,
we will discuss how these goals are achieved. A naive implementation
of a hash table will immediately issue update requests to the hash
table as the data points are received. The vast majority of the
write operations will be random page level writes due to the lack of
locality, which is inherent in hash function design.  As mentioned
before, the cost of such operations will also increase the cost of
cleans and random writes.

A desirable property for a hash table would be  {\em block-level
updates} and {\em semi-random writes}. The {\em block-level update}
refers to the case when there are multiple updates written to a
block, and they are all accomplished at one time. If there are $k$
updates written to a block, we should combine them into one
block-level write operation. This can reduce the number of cleans
from $k$ to one. The semi-random writes refer to the fact that the
updates to a  particular block are in the same order as they are
arranged on the block, even though updates to different blocks may
be interleaved with one another.

We  give an example of semi-random writes. If we consider the pair
$(Blockid, pageid)$ sequence $(3, 2)$, $(2, 1)$, $(4, 3)$, $(2, 4)$,
$(2, 5)$, $(3, 4)$, $(4, 6)$, $(3, 5)$, $(3, 7)$, this is considered
semi-random, because the page updates to a {\em particular block}
are arranged sequentially by their order of page id. Recall that
sequential write patterns improve latency \cite{nath}. This is
because of how the flash translation layer works.  The existing
methods in the flash translation layer are typically lazy;  when the
$i$th logical page of a block is written, the FTL copies and writes
{\em only} the first $i$ pages to a newly allocated block, instead
of all the pages.  Later, when page $j
> i$ is modified, {\em only} the pages $(i + 1)$ to $j$ are moved and written to the new
block.  Note that this would not be possible if $j$ were less than
$i$.  The semi-random writes would improve the write latency of an
SSD because SSD write performance improves under sequential pattern
of writes \cite{nath}.   Thus, the sequential ordering is useful in
minimizing the unnecessary copying of pages from old blocks to new
blocks.

\subsection{Hash table designs}
A variety of low level structures can be maintained in order to
accomplish the desirable properties discussed earlier. We will
design a number of such  hash table maintenance  schemes. All of
these schemes use a combination of these low level structures in
different ways. However, we would like to introduce these low level
structures at this stage in order to ease further discussion. Recall
that we combine an open hash table in main memory with a closed-hash
table on the SSD. This open (or secondary) hash table is typically
implemented in the form of a RAM buffer denoted as  $H_R$. The RAM
buffer will contain updates for each block of the SSD and execute
batch updates to the primary hash table on disk,  or {\em data
segment} (denoted by $H_D$), at the block level. This approach can
reduce block level cleaning operations.

\subsection{Memory Bounded Buffering}
\begin{figure}[hbtp]
\centerline{
\includegraphics[width=2.8in, height=1.3in]{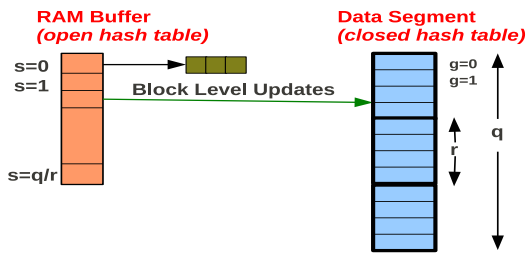}
}
\caption{Hash Table with RAM buffer}
\label{fig:default1}
\end{figure}

The  overall structure of the common characteristics of the  hash
table architectures presented in this paper is illustrated in Figure
\ref{fig:default1}.  We refer to this scheme as {\em Memory Bounded
Buffering} or {\em MB}. The RAM buffer in the diagram is an open (or
secondary) hash table and the data segment is a closed  (or primary)
hash table. There are $s$ slots, each of which corresponds to a {\em
block} in the data segment. The maximum capacity of the data segment
is $q$ pages, $r$ pages per block and $g$ entries per page. Thus,
the number of slots in the secondary hash table, $s$, must be equal
to $q/r$. Updates are flushed onto the SSD one block at a time.
Because of the relationships between the hash functions of the
primary and secondary table, the {\em merge} process of a given list
requires access to only a particular set of SSD blocks which can be
maintained in main memory during the merging process.  This may of
course involve the insertion of new items that are not present in
the data segment and items that collide with entries already inside
of the data segment.

\subsection{Memory and Disk Bounded Buffering}
\begin{figure}[hbtp]
\centerline{
\includegraphics[width=2.8in, height=1.3in]{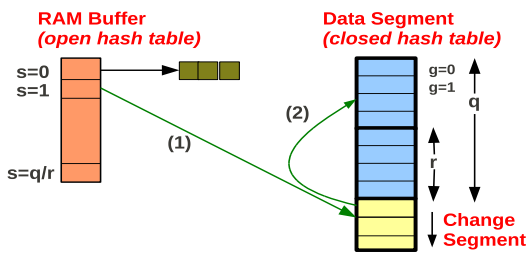}
}
\caption{Hash Table with RAM buffer and Change Segment}
\label{fig:default2}
\end{figure}

Since  {\em $H_R$} is  main-memory resident, it is typically
restricted in size. Therefore,  a second buffer can be implemented
on the SSD itself.  This new segment is referred to as the  {\em
change segment} or $S_E$.  The change segment acts as a second level
buffer. When {\em $H_R$} exceeds its size limitations, the contents
are sequentially written to the change segment at the page level
starting from the first available page in an operation known as {\em
staging}. When full, the change segment {\em merges} with the data
segment and begins from the top of the change segment. A page in the
change segment may contain updates from multiple blocks because
pages are are packed with up to $g$ entries irrespective of their
slot origin.  Thus, the change segment is organized as a log
structure that contains the flushed updates of the RAM buffer. This
takes advantage of the semi-sequential write performance of the SSD
and increases the lifetime of the SSD.  The space allocated to the
change segment is in addition to the space allocated to the data
segment.  This hash table (with change-segment included) is
illustrated in Figure \ref{fig:default2}.  It is important to note
that a {\em stage()} operation differs from a {\em merge()}
operation in two ways, Specifically, stages write at the page level
while merges operate at the block level.   Furthermore, stages
involve updates to the change segment while merges involve updates
to the data segment.

There are two types of architectures for the change segment.
In the first design, the change segment $S_E$ is viewed as a collection of blocks where
each block holds updates from multiple lists from $H_R$. In other
words, multiple blocks in the data segment are mapped to a single
block in the change segment.  We arrange the change segment in a way
such that each change segment block holds the updates for $k$ data
segment blocks. The value of $k$ is constant for a particular
instantiation of the hash-table, and can be determined in an
application-specific way. For an update-intensive application, it is
advisable to set $k$ to a smaller value at the expense of SSD space.

When a particular change segment block is full, we merge the
information in the change segment  to the data segment blocks. The
advantage of this method is best demonstrated if the RAM buffer is small.
In that case, it will cause frequent merges onto the data segment
under the {\em MB} design of the hash table. By adding the change
segment, we are providing a more efficient buffering mechanism.  Staging a segment is more
efficient than merging it because the change segment is
written onto the SSD with a straightforward sequential write, which is
known to be efficient for SSD.  This approach is called {\em Memory Disk Buffering} or {\em MDB}.

In this variation of the {\em MDB} scheme, (which we henceforth will
refer to as {\em MDB-L} for {\em MDB-Linear}) the space allocated
for the change segment is viewed as a single large monolithic chunk
of memory without any subdivisions. This view resembles a large log
file. Thus,  the change segment blocks are not assigned to $k$ data
segment blocks. The writes to the change segment are  executed in
FCFS fashion.  This type of structure mimics a log-structured file
system and fully takes advantage of the SSD strength in sequential
writes.  We maintain a collection of pointers to identify the
ranges, measured in pages, that a particular slot in the RAM buffer
has been staged. These pointers are similar to the indexing
information~\cite{rosenblum1992design} maintained in log-structured
file systems that helps in reading the files from the log
efficiently.

A merge operation is triggered when the change segment is full. The
collection of pointers can be used to identify the pages a
particular block was staged.  This process produces random reads on
the change segment because the ranges span multiple stage points.
The reads are also repetitive because a page may contain entries
from multiple blocks because of the staging process.  During a
stage, entries from multiple blocks may be packed into a single
page.  During a merge, each page will be requested by each data
segment block that has entries staged onto it.  After all of the
pages for a particular data segment block are read from the change
segment, the entries are merged with the corresponding data segment
block.

\subsection{Element Insertion  and Update Process}
The element insertion process is designed to perform individual
updates on the memory-resident table only, since this can be done in
an efficient way.  Such changes are later rolled on to the RAM
buffer (which is in turn rolled on to the change segment for some of
the schemes).

For each incoming record $x$, we first apply the hash function
$s(x)$ in order to determine the slot to which the corresponding
entry belongs. We then determine if the key $x$ is present inside
the corresponding slot $s(x)$. If the element is found, then we
increase its frequency.  The second case is when the key $x$
is not contained inside the buffer which is pointed to by the slot
$s(x)$. In such a case, we add the key $x$ as a new element to the
RAM buffer. The size of the RAM buffer increases in this case.
If the RAM buffer has grown too large, it is flushed either directly
 onto the change segment or the SSD itself,
depending upon whether or not the change segment is implemented in
the corresponding scheme. Because of the relationship between the
hash functions of the RAM buffer and the SSD based hash table, such
an update process tends to preserve the locality of the update
process, and if desired, can also be made to preserve semi-random
write properties.

During the insertion process of new items, linear probing may occur
because $H_D$ is a closed hash table.  If the linear probing process reaches the end of the current
SSD block, then we {\em do not} move the probe onto the next block.
Rather,  an overflow region is allocated within the SSD table which
takes care of probing overflows beyond block boundaries. The last
index of the last page of an SSD block becomes a pointer assigned to
the overflow region. The entry that was resident at this position
now resides in the conflict region alongside the newly inserted
entry. Thus,  the {\em data segment} is a collection of blocks with
logical extensions.  The overflow region, a collection of SSD blocks,
 is allocated when
the hash table is created.  Its size can vary depending on user
specifications.  The blocks are written one at a time and the pages
are assigned when needed.  When a overflow region is needed, a page
is assigned.  If another region is needed, another page is allocated
and the previous page points to the newly allocated page. When a
block is full, another block is used.  An overflow block may contain
entries from multiple blocks in the SSD data segment $H_D$
because the regions are allocated a page at a time when they are needed.

\subsection{Implementing Deletion Operations}
 It is also possible to implement deletion operations in the hash
 table. There are two kinds of deletion possible:
\begin{itemize}
\item A deletion operation reduces the frequency of an item by 1.
This case is trivially solvable by using the approach  for
insertion, except that we use a frequency of -1 for the incoming
element. Entries with frequency value of 0 are not retained in the
memory-resident table, and entries with negative frequencies are
allowed. These negative frequencies are appropriately transferred to
the drive-resident table during batch updates.
\item A deletion operation completely removes an element from the
hash table irrespective of its frequency. This case is more complex,
and is discussed below.
\end{itemize}

If an item is deleted from the data segment, it can either be
removed or its frequency can be set to zero.  If it is removed,
there will be added complexity to queries, updates, and inserts.
This is because of the way in which the linear probing process is
implemented.  During a query, if an empty slot is encountered during
the probing process, the query terminates (with the guarantee that
the item is not found anywhere else in the hash table) because of
this contiguity assumption.  However, the removal of entries in a
deletion process can violate this contiguity assumption. This is
because the empty slot encountered during linear probing may be the
result of a previously filled slot, which was removed by a deletion
process.  In such a case, the desired  entry may reside  beyond the
empty slot, and it may no longer be possible to terminate after the
first instance of an empty slot. This could potentially invalidate
the correctness of the query processing.

This problem can be handled during the {\em merge} of a block or
periodically.  In both cases, the block is loaded into main  memory.
The entries are hashed inside  a main memory block, but the deleted
items are ignored during this process. This will ensure that the
entries are contiguous.  The newly hashed entries are then
re-written directly to the data segment block.

\subsection{Query Processing}
In the simple hash table, queries are fulfilled by an I/O request to
the data segment.  However, in our proposed designs the corresponding entry
may be found either in the change segment or the RAM buffer.
Therefore, the  query processing approach must search the change
segment and the RAM buffer in addition to the data segment.  Thus, the frequency of a queried item is the total frequency found in the change segment, RAM buffer, and data segment. The
search of the RAM buffer may be inexpensive because it is in main
memory. On the other hand, access to  the change segment requires
access to the SSD.

For the case of the data segment, the query processing approach is
quite similar to that of standard hash tables. A hash function is
applied to the queried entry in order to determine its page level
location inside of the data segment.  If the entry is found, the
frequency is returned.  If the item is not found, linear probing
begins because the disk hash table is a closed hash.  Linear probing
halts if the entry is found or an empty entry was discovered. The
query processing of  the change segment requires locating the entry.
The location of the entry may reside in multiple segments due to
repeated flushing of the RAM buffer.

Recall that {\em MDB} partitions the change segment.  When a RAM
bucket is staged, it is always written to the same change segment
block.  We locate the appropriate change segment block and bring it
into memory to be searched.  In {\em MDB-L}, RAM buckets can reside
on multiple pages, and  thus we must issue random page reads.  We
expect {\em MDB-L} to be faster because of  page level access.

\section{Experiments}
\label{sec:experiments}

In this section we present an empirical analysis of the hash table designs
discussed in the previous sections.  We evaluate the performance of the three main schemes discussed in this article, namely {\em MB}, {\em MDB} and {\em MDB-L}. Broadly, our objectives are to understand the I/O overheads of various schemes and their query performance.
Additionally, since SSD disks permit a limited number of clean operations, it is also important to
quantify the wear rate of the devices.  We begin with a discussion of the experimental setup.
\subsection{Experimental Setup}

To evaluate our hash table configurations, we used the DiskSim
simulation environment \cite{disksim}, managed by Carnegie Mellon
University; and the SSD Extension for this environment created by
Microsoft Research \cite{ssdAdOn}. We used the Disksim simulator
with an SSD extension by Microsoft Research.  We operated Disksim in
slave mode.  Slave mode allows programmers to incorporate Disksim
into another program for increased timing accuracy.

We ran our experiments on three different configurations of the latest representative NAND flash SSDs from Intel (see Table\ref{tab:ssd-conf} for details).
Among these, two SSDs are MLC (Multi-Level Cell) and the other is SLC (Single-Level Cell)
based SSDs. We have chosen from both MLC and SLC because of their differing
characteristics. While MLCs provide much higher data density and lower cost (which makes it
more popular), it has a shorter lifespan and slow read/write performances. SLC,
on the other hand, has faster read/write performances and a significantly longer
lifespan. SLCs also entail lower internal error rate making them preferable for
higher performance, high-reliability devices \cite{hpPaper}.

All hash table experiments involve inserting, deleting, and updating
key value pairs.  The size of the RAM buffer is parameterized on the
size of the data segment and expressed as a percentage The rationale
here is that we believe that an end application may need to create
multiple hash tables on the same SSD. Moreover, the characteristics
of access may vary across applications (i.e. one may want different
RAM buffer sizes for each hash table).  The change segment is
likewise parameterized and the overflow segment for all experiments
was set to a minimal value (one block) since this was found to be
sufficient.  Key-value pairs are integer pairs.

We conducted our experiments on a DELL Precision T1500 with an Intel
\textregistered \ Core \texttrademark  \ i7 CPU 860@2.8GHz with 8Gbs
of memory and 8 cores running Ubuntu 10.04. Our code was implemented
in C++.  The  $H_R$  data structure utilized the C++ Standard
Template Library \cite{cplus} for its implementation.  The RAM
buffer buckets that correspond to data segment blocks are arranged
inside a C++ vector and their indexes correspond to their placement
on the data segment. For example, the first block inside the data
segment corresponds to the first block in the RAM buffer.  The data
segment can be viewed as an array logically divided into blocks and
further divided into pages. 

\subsection{TF-IDF}
To demonstrate the efficacy of our methods, we implemented the TF-IDF algorithm, see Section \ref{sec:Intro},
using our hash table designs.  In our experiments, we compute the TF-IDF score of all words in our corpus.  Our hash table contains the frequencies of each keyword.  As we read in each document, we compute the frequency of each word and store it in our counting hash table.

\begin{table}[!t]
\centering

    \begin{small}
    \begin{tabular}{|c|c|c|c|}
    \hline
            &   \multicolumn{3} {c|} {\bfseries SSD Configurations} \\
    \cline{2-4}
            &   {\bfseries  MLC-1}  &   {\bfseries MLC-2}   &   {\bfseries SLC}           \\
    \hline
    \hline
    Capacity    &   40GB        &   80GB        &   32GB    \\
%       &   and     &           &   and \\
%       &   160GB       &           &   64 GB   \\
    \hline
    Flash Memory    &   MLC &   MLC     &   SLC \\
    \hline
    Page Size (KB)  &   4   &   4       &   4   \\
    \hline
    Sustained   &   Upto 170    &   Upto 250    &   Upto 250    \\
    Sequential  &           &           &   \\
    Read (MB/s) &           &           &   \\
    \hline
    Sustained   &   Upto 35     &   Upto 70     &   Upto 170    \\
    Sequential  &           &           &   \\
    Write (MB/s)&           &           &   \\
    \hline
    Read    &   65      &   65      &   75  \\
    Latency ($\mu$s)    &       &           &   \\
    \hline
    Write   &   110      &   85     &   85  \\
    Latency ($\mu$s)&           &           &       \\
    \hline
    Cost (USD)  &   109.99      &   224.99      &   377.79  \\
    \hline
    \hline
    \end{tabular}
    \end{small}
    \caption{SSD Configurations}     %
    \label{tab:ssd-conf}
\end{table}

\subsection{Data Sets}
We use two datasets: a Wikipedia and MemeTracker dump, which
are essentially large \emph{text} files.

\noindent{\bf Wiki:} The first data set we use is a collection of
randomly collected Wikipedia articles.  We chose $100,000$ random
wikipedia articles collected from Wikipedia's publicly available
dump\footnote{See http://dumps.wikimedia.org/}. Our $100,000$ random
articles were approximately $1GB$ in size. This dataset contains
$136,749,203$ tokens (keywords) with $9,773,143$ unique entries. For
the testing of this data set, our hash table was set to $100MB$. On
a $128$ page per block SSD this amounts to approximately $205$ SSD
blocks allocated to the data segment.

To evaluate I/O performance during inserts or updates\footnote{ As
noted earlier deletes are handled as inserts with a negative count.}
we simply inserted (or update) tokens (corresponding counts) into
the hash table. Statistics and times for various operations (cleans,
merges, stages etc.) are computed and discussed shortly.

To evaluate query performance we first processed 35 million
tokens. Subsequently, roughly 100 million words were inserted.
Simultaneously with inserts, we also issued a million queries
interleaved randomly across inserts.  A query is a hash table lookup.  In the TF-IDF context, this corresponds to 
{\em "how frequent is a keyword"} which allows us to compute the TF-IDF score of a keyword.  
Of these queried items, on average (spread across 10 different random workloads) $933,139$ of
them were present  inside the hash table at query time.

\noindent{\bf Meme:} The second data set we report is the
MemeTracker dataset\footnote{See http://memetracker.org/}. We
downloaded the \textit{August 2008} dataset. We found $17,005,975$
unique entries and $402,005,270$ total entries. Since this dataset
is slightly larger our hash table size was $200MB$. On a $128$ page
per block SSD this translates to approximately $410$ SSD blocks
allocated to the data segment.

I/O performance is evaluated similar to how we handle things for the
Wiki dataset. For query performance, the first 130 million words
were inserted into the hash table. Subsequently, the remaining 270
million words were interleaved with about one million queries. Of
these queried items, (spread across 10 random workloads) $959,731$
of them were found inside the hash table.

\begin{figure*}[htp]
    \centering
    \begin{subfigure}[b]{0.33\textwidth}
    \centering
    \includegraphics[width=\textwidth]{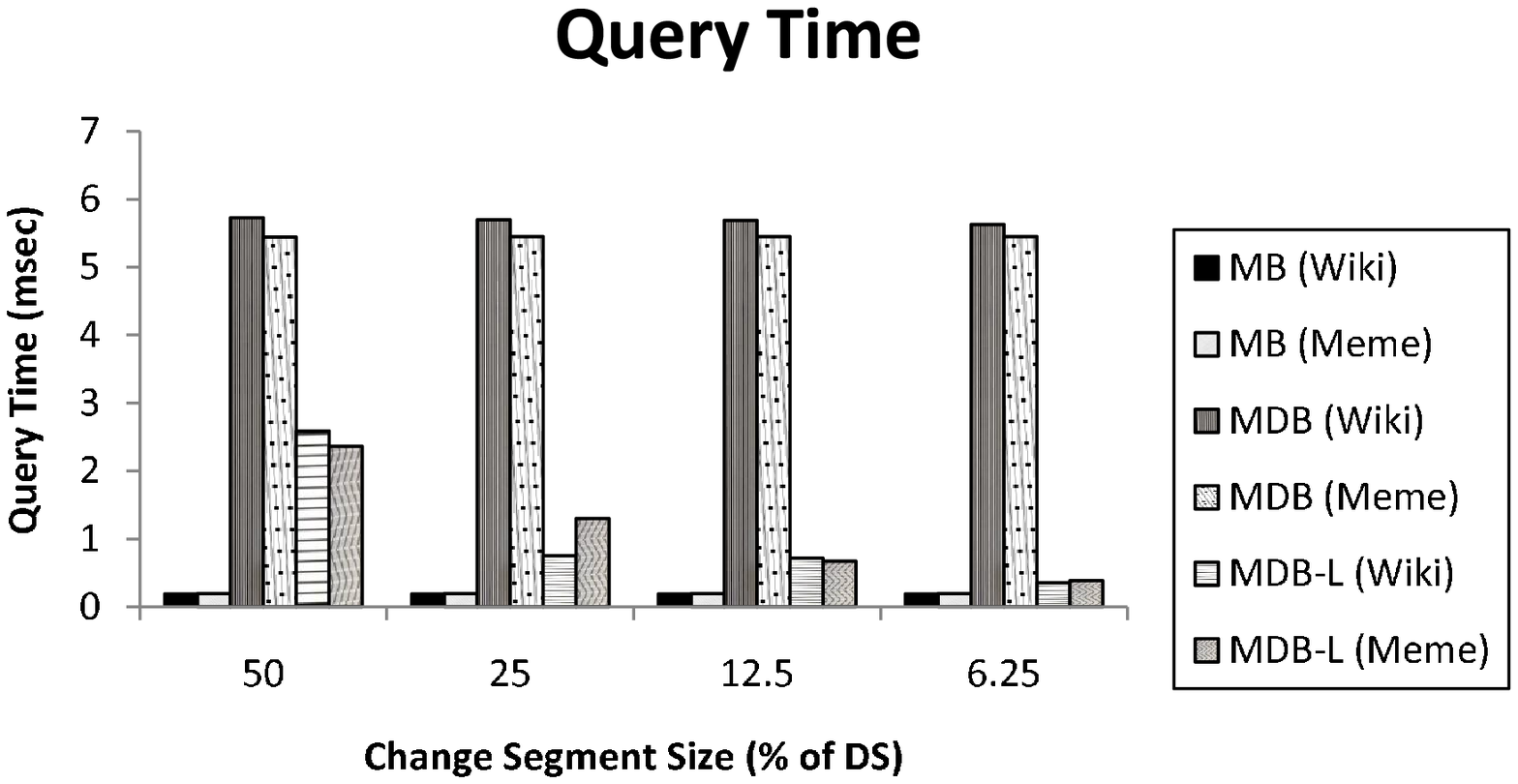}
    \caption{Query times for varying change segment sizes}
    \label{fig:qtK}
    \end{subfigure}
    \begin{subfigure}[b]{0.33\textwidth}
    \centering
    \includegraphics[width=\textwidth]{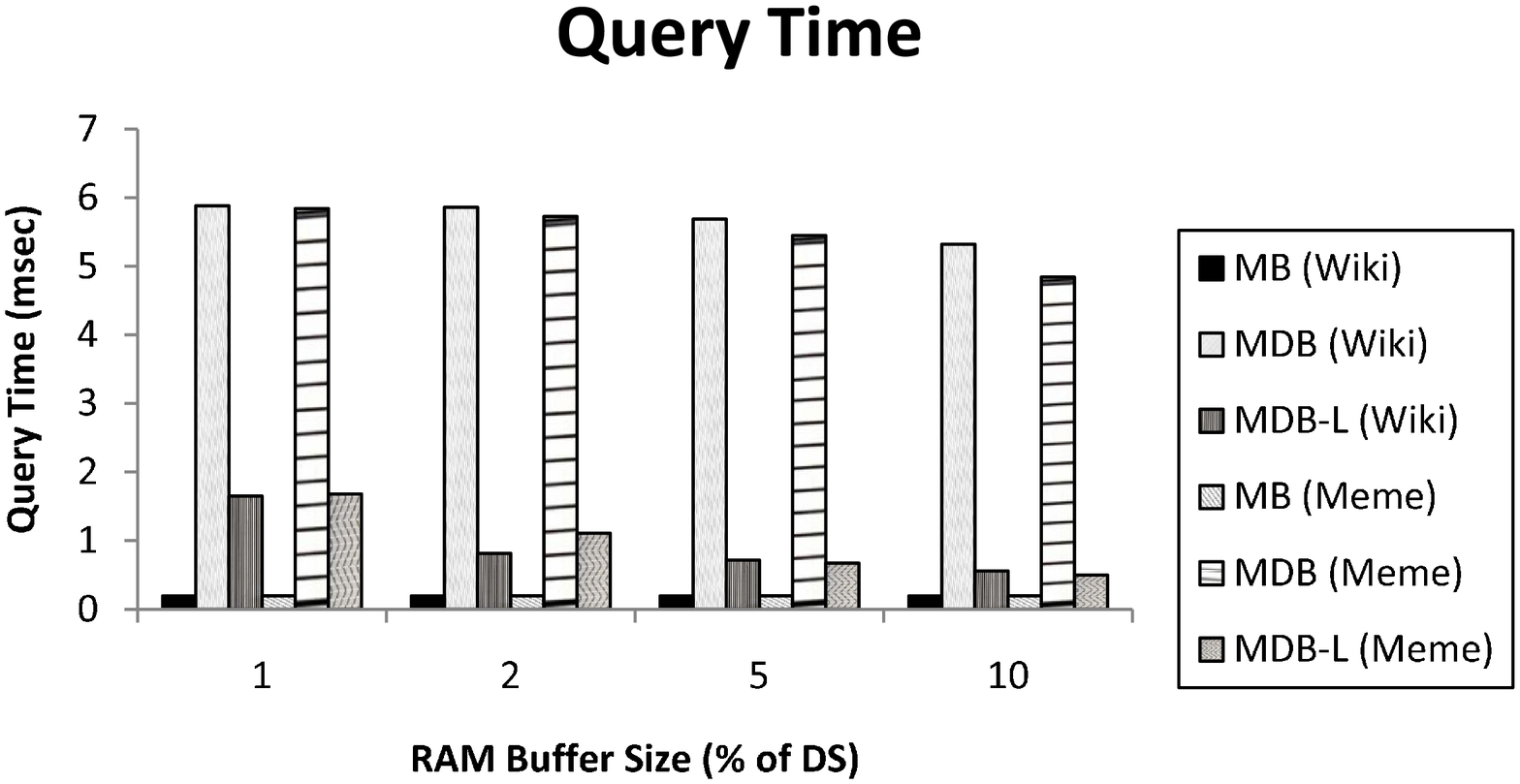}
    \caption{Query times for varying RAM buffer}
    \label{fig:qtRAM}
    \end{subfigure}   
    \begin{subfigure}[b]{0.33\textwidth}
    \centering
    \includegraphics[width=\textwidth]{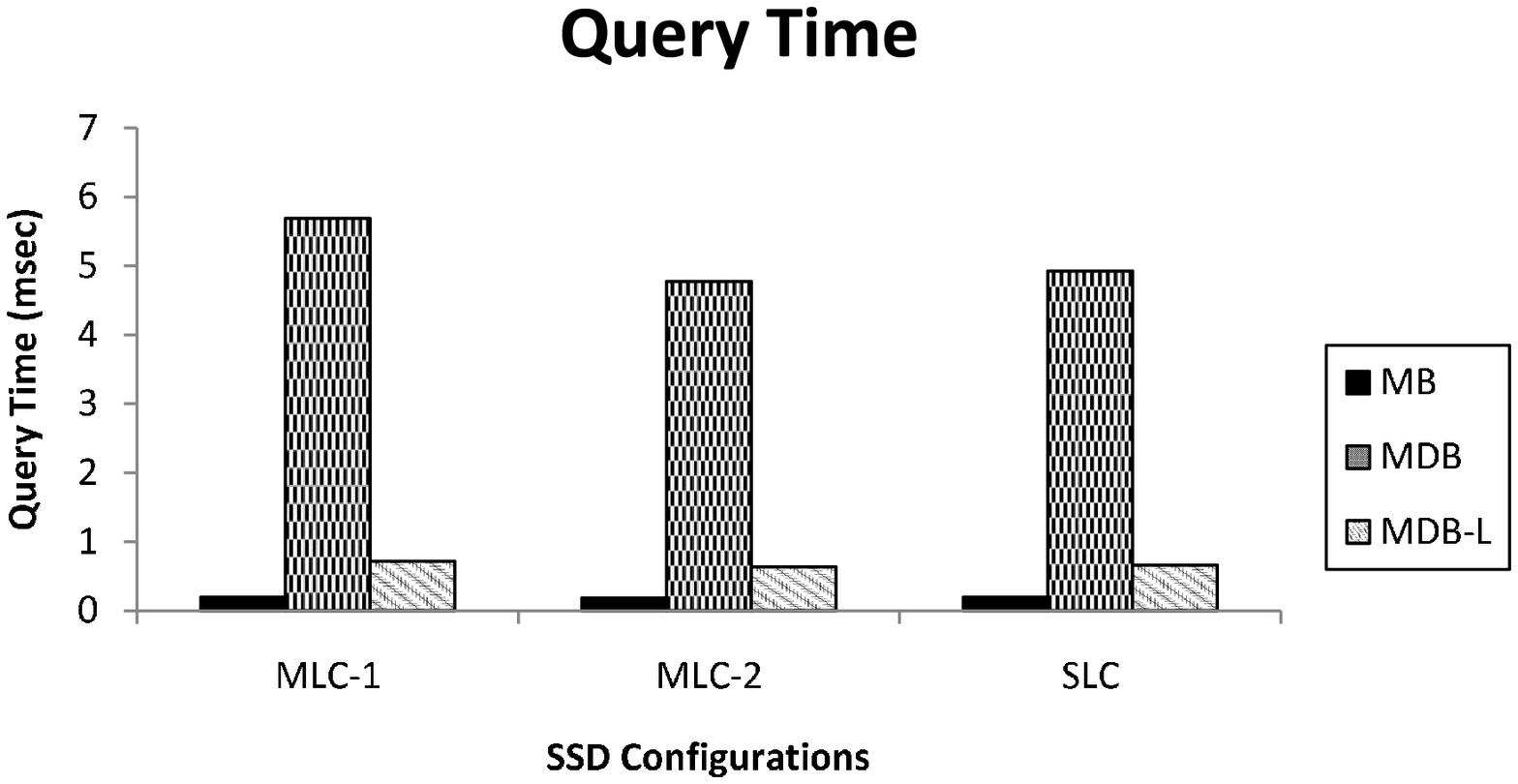}
    \caption{Query times for all SSD configurations}
    \label{fig:qtLMH}
    \end{subfigure}    
    %\subfigure[Query times for varying change segment sizes]{\label{fig:qtK}\includegraphics[width=3in]{qtK.eps}
    %\subfigure[Query times for varying RAM buffer]{\label{fig:qtRAM}\includegraphics[width=3]{qtRAM.eps}
    %\subfigure[Query times for all SSD configurations]{\label{fig:qtLMH}\includegraphics[width=3in]{qtLMH.eps}
  \caption{Query Times}
  \label{fig:qtimes}
\end{figure*}

\begin{figure*}[htp]
   \centering
   \begin{subfigure}[b]{0.49\textwidth}
    \centering
    \includegraphics[width=\textwidth]{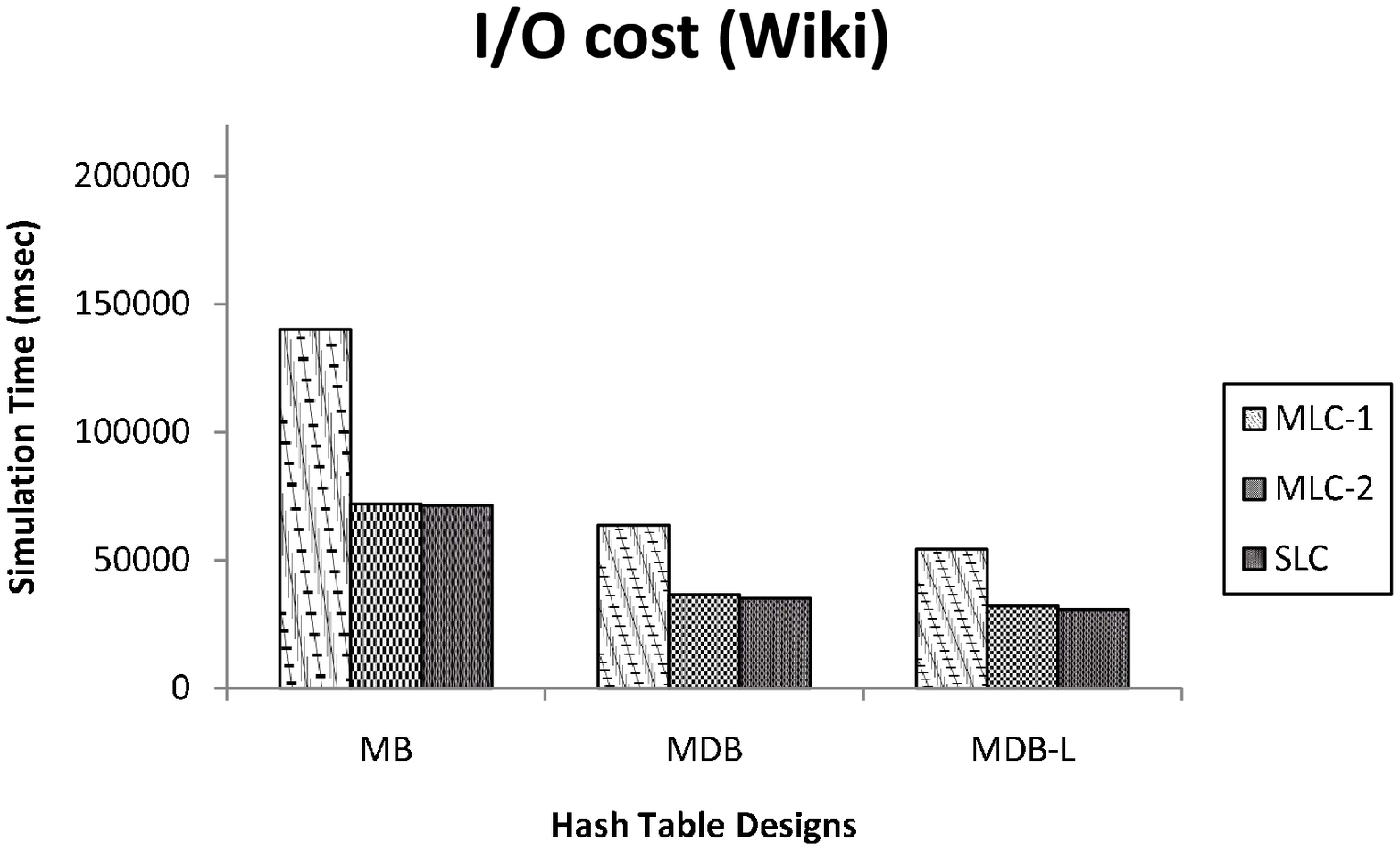}
    \caption{I/O cost for {\em Wiki} dataset}
    \label{fig:iowiki}
    \end{subfigure}
    \begin{subfigure}[b]{0.49\textwidth}
    \centering
    \includegraphics[width=\textwidth]{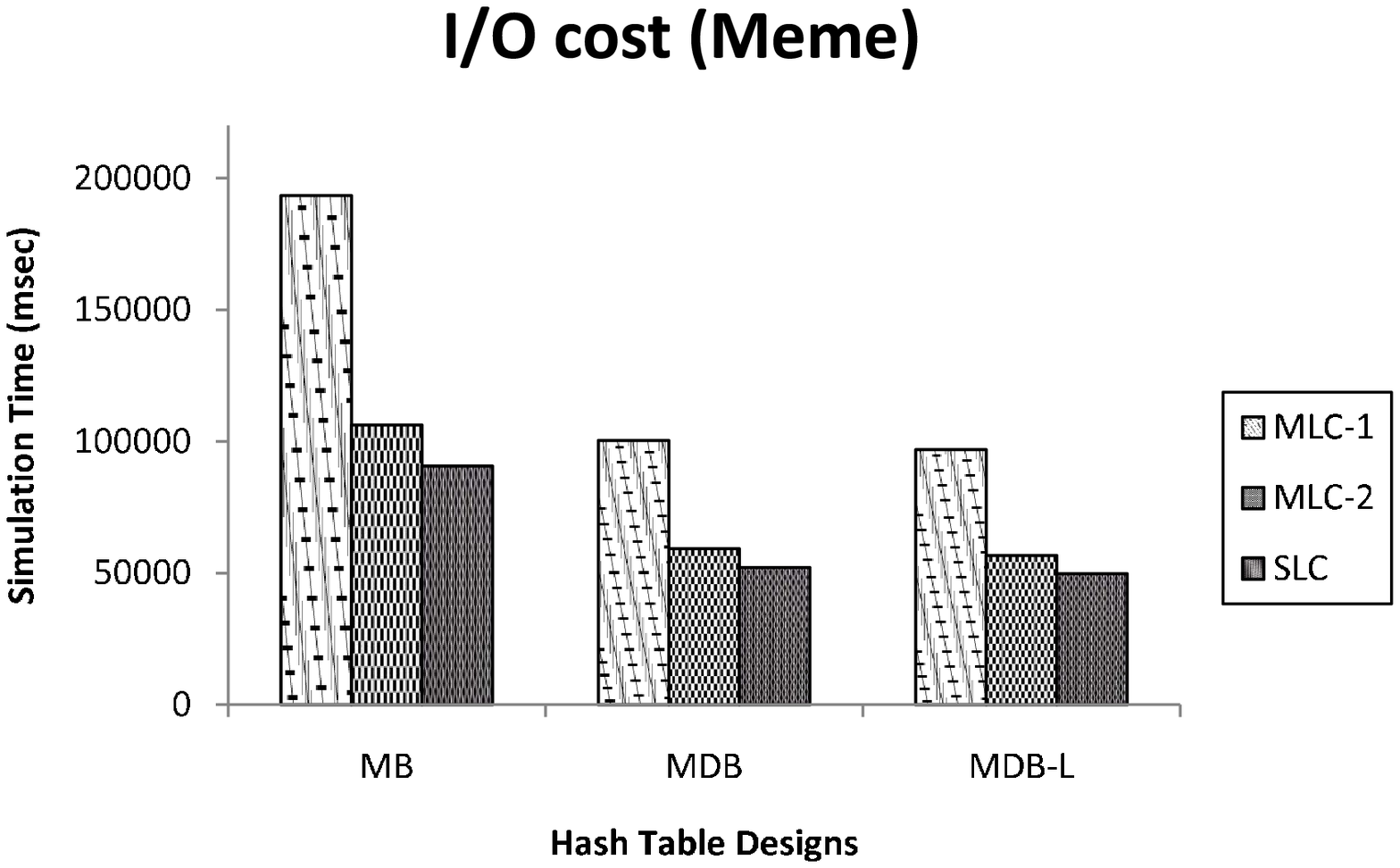}
    \caption{I/O for {\em Meme} dataset}
    \label{fig:iomeme}
    \end{subfigure}  
   %\subfigure[I/O cost for {\em Wiki} dataset]{\label{fig:iowiki}\includegraphics[width=3in]{iowiki.eps}
   %\subfigure[I/O for {\em Meme} dataset]{\label{fig:iomeme}\includegraphics[width=3in]{iomeme.eps}
  \caption{I/O Costs}
  \label{fig:iocost}
\end{figure*}

\begin{figure*}
   \centering
   \begin{subfigure}[b]{0.3\textwidth}
    \centering
    \includegraphics[width=\textwidth]{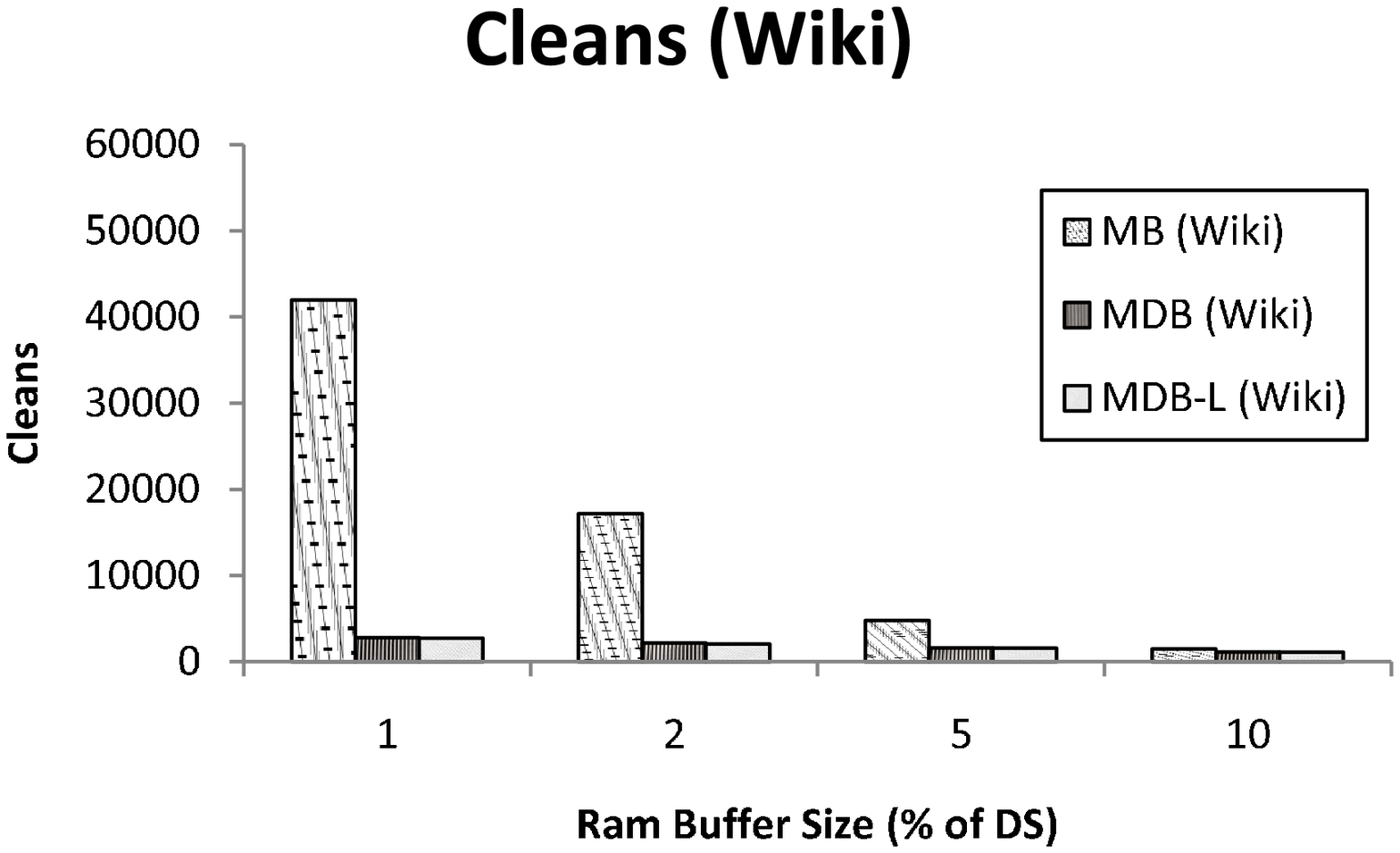}
    \caption{Cleans for {\em Wiki} dataset}
    \label{fig:cleans-wiki}
    \end{subfigure}
    \begin{subfigure}[b]{0.3\textwidth}
    \centering
    \includegraphics[width=\textwidth]{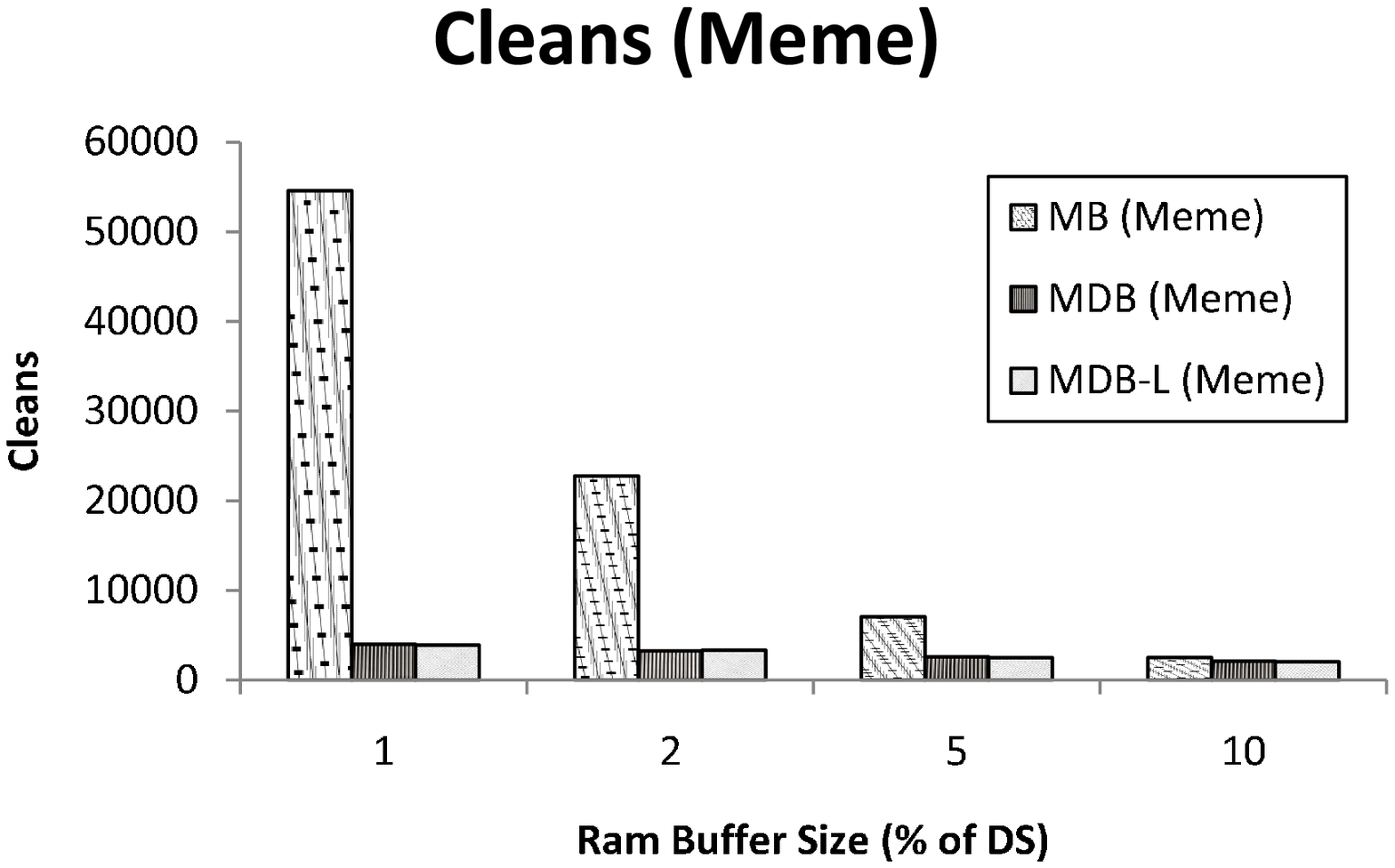}
    \caption{Cleans for {\em Meme} dataset}
    \label{fig:cleans-meme}
    \end{subfigure}
    \begin{subfigure}[b]{0.3\textwidth}
    \centering
    \includegraphics[width=\textwidth]{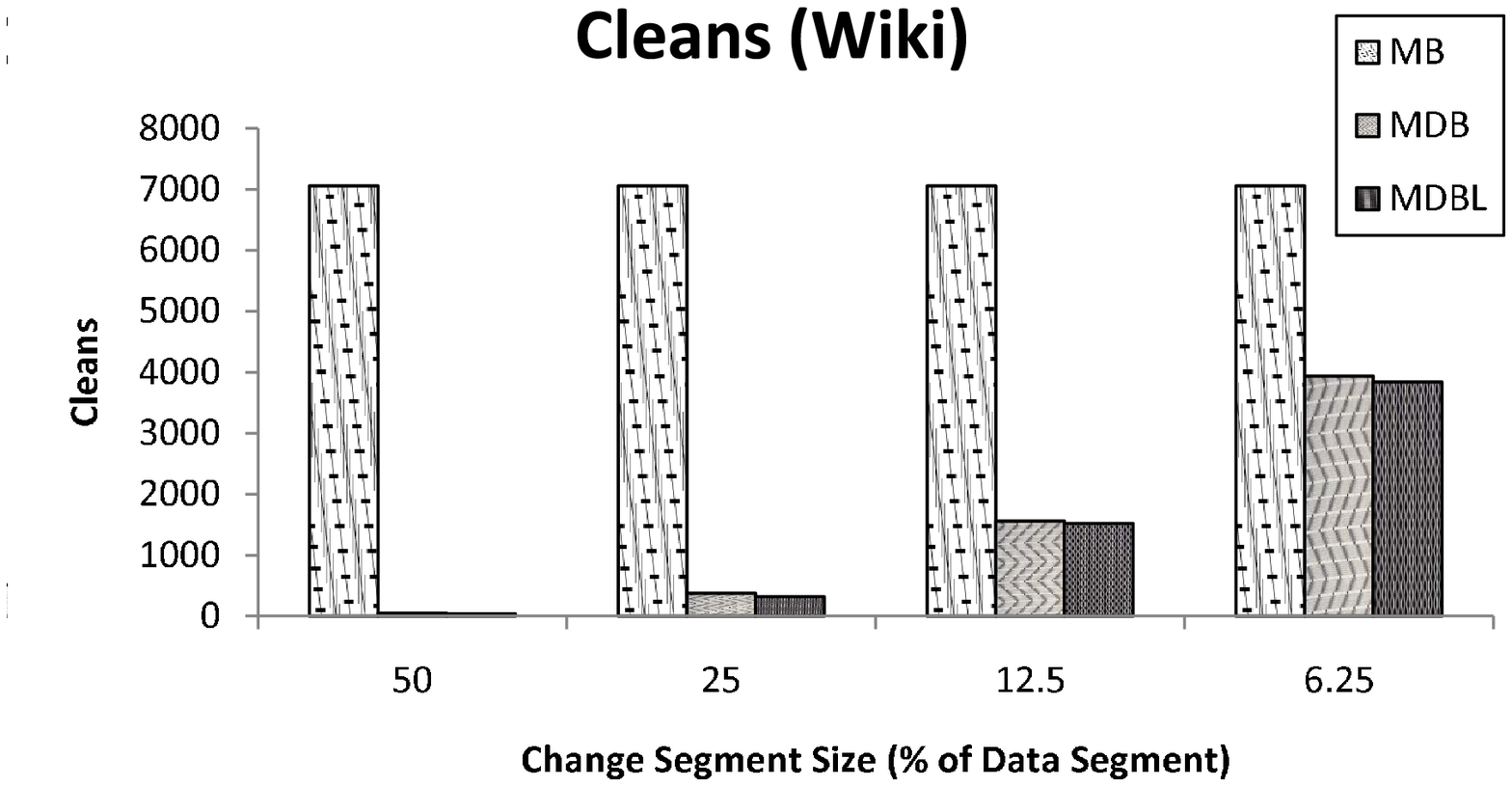}
    \caption{Cleans for varying change segment sizes}
    \label{fig:cleansK}
    \end{subfigure}
   %\subfigure[Cleans for {\em Wiki} dataset]{\label{fig:cleans-wiki}\includegraphics{width=2.4in]{cleans-wiki.eps}
   %\subfigure[Cleans for {\em Meme} dataset]{\label{fig:cleans-meme}\includegraphics{width=2.4in]{cleans-meme.eps}
   %\subfigure[Cleans for varying change segment sizes]{\label{fig:cleansK}\includegraphics{width=2.4in]{cleansK.eps}
  \caption{Cleans}
  \label{fig:cleans}
\end{figure*}

\subsection{Query Time Performance}

In all our graphs, the Y-axis is the average time per query in milliseconds.
Results on both Wiki and Meme are provided in Figure~\ref{fig:qtimes}. The main
trends we observe include: i) the query time for {\em MB} are quite low (does
not have a change segment);
ii) the query time for {\em MDB} is quite high and does not drop significantly
with reduction in change segment size;
iii) the query time for {\em MDB-L} improves dramatically
with a reduction in the change segment size; and iv) query times
for Meme are marginally lower than the query times for Wiki for both
{\em MDB} and {\em MDB-L} .
These trends can be explained as follows. Query costs for MB are essentially
fixed since they essentially have to combine the counts from the memory buffer
(negligible) and require typically a page read to access the requisite
information from the data segment.
Query costs for {\em MDB}
require consolidation of information from the memory buffer (negligible)
and from the change segment (expensive -- dominated by block level reads)
and the data segment (usually a single page read).
Query costs for {\em MDB-L} require consolidation across the memory buffer
(negligible), the change segment (typically requiring a few page reads
which are significantly reduced as the size of the change segment is reduced)
and the data segment (usually a single page read). This is reflected in our first experiment, shown in Figure~\ref{fig:qtK} for MLC-1,  in which we varied the change segment while fixing the RAM buffer to 5\%.
With regards to the difference between Wiki and Meme query times,
upon drilling down into the data, we find that on average there are 11.5\%
more page reads for Wiki.
This may be an artifact of the linear probing
costs within both datasets,
given the fact that the ratio of number of unique tokens to hash table
size is slightly higher for Wiki.

For the second experiment, again on the MLC-1 configuration,
we fixed the change segment to 12.5\% and varied the the RAM buffer for
both datasets(see Figure~\ref{fig:qtRAM}).
We observe that with an increase in RAM buffer
size that : i) {\em MB}
shows a negligible change
in average query time;
ii) {\em MDB} shows decrease in average query time;
iii) {\em MDB-L} shows a significant decrease in average query time performance;
and iv) query times on Meme are typically faster than those on Wiki.

To explain these trends we should first note that increasing RAM buffer
size has the general effect of reducing the number of stage operations,
and thus the average size of the amount of useful information within
the change segment.
Thus the time it takes to consolidate
the information within the change segment in order to answer the query,
is on average lower, for both {\em MDB} and {\em MDB-L}.
For {\em MDB-L} the improvement is more marked because fewer page reads
are required. The explanation for why query times are lower for Meme
are similar to what we observed for the previous experiment.

The third experiment we performed on query time performance was to evaluate
the performance of the three SSD configurations on the Wiki dataset shown in Figure~\ref{fig:qtLMH}.
Here the RAM buffer was set to 5\% and the Change Segment was set to
12.5\%. The results are along expected lines in that average query times
are slightly better on MLC-2 and SLC over MLC-1 for the {\em MDB} method. The superior read performance for both page level and block level operations are the primary reason. This difference
is marked in the case of {\em MDB} but for both {\em MDB-L} and {\em MB} the difference
is negligible.  {\em MDB} requires a block level read for a query and the performance difference for this type of operation is more pronounced for MLC-2 and SLC, over  when compared with MLC-1.

To conclude we should reiterate that the query performance times we observe
here are for our update-intensive
query workload where we interleaved queries with inserts (averaged over multiple runs). In this environment
the query time performance of MB is always the fastest.
For {\it reasonable parametric settings MDB-L
typically approaches MB in performance while MDB
is always an order of magnitude worse in terms of performance}.
We should note that we also evaluated
query times for all three methods
in more stable settings (few updates/inserts)
In such a {\it stable setting we found that the query times for all
three methods
was identical}. The query cost essentially boils down to a
page read or two on the data segment (since the change segment is
empty and does not factor).
Furthermore, MDB is bounded by a single block read while the query time of MDB-L may vary.  However, 
as our results indicate, the pointer guided page level accesses of MDB-L provide efficient read access that outperforms MDB.

\subsection{I/O Performance}

In this section we examine the I/O performance of the three strategies.
To ignore the impact of queries in this section,
our workloads for both datasets simply insert all the tokens or words
into their respective hash tables.
In our first experiment, we report overall I/O cost from the perspective of
the SSD for the three SSD configurations for both
Wiki (see Figure~\ref{fig:iowiki}) and Meme (see Figure~\ref{fig:iomeme}).
The RAM buffer is set to 5\% and the change segment is set to 12.5\%
in this experiment.

The main trend we observe are that both {\em MDB-L} and {\em MDB} require comparable
yet significantly lower I/O costs than {\em MB}. This is primarily attributable
to the presence of the change segment which enables {\it sequential (MDB-L)
or semi-random writes (MDB)}. Additionally, as we shall see shortly,  {\em MB} requires a large number of erasures which also contribute to the overall I/O cost.
Another trend we observe is that among the
SSD configurations SLC and MLC-2 offer comparable performance with
a slight edge to SLC. MLC-1 is quite a bit slower.
This is primarily attributable to the superior write bandwidth of SLC and MLC-2. Finally we observe that the overall I/O times are higher for Meme over Wiki (larger dataset and larger hash table).

Not shown in our reports are the performance measures for a hash
table without the use of a buffer.  The advantage of this scheme is
fast query times because queries are only page level reads on the
data segment.  However, results show that such a hash table would
induce 1,680,323 cleans for the Wiki dataset and 6,669,932 cleans
for the Meme dataset.  The I/O performance are on the order of 615
times slower on the Wiki dataset and 1500 times slower for the Meme
dataset for reported times for the results in Figure
\ref{fig:iocost}.  This increase is caused by cleaning time and
random page writes.  It is clear that there is a benefit from our
designs.

Next we discuss the  breakdown of page and block level operations,
merge and stage operations.
\begin{table*}[!t]
\centering

    \begin{small}
    \begin{tabular}{|c|c|c|c||c|c|c|c||c|c|c|c|}
%    \hline
    \hline
    {\bfseries Change}  &   {\bfseries RAM} &   \multicolumn{10} {c|} {\bfseries Methods} \\
    \cline{3-12}
    {\bfseries Segment} &   {\bfseries Buffer}  &   \multicolumn{2} {c||} {\bfseries MB}    &   \multicolumn{4} {c||} {\bfseries MDB}   &   \multicolumn{4} {c|} {\bfseries MDB-L} \\
    \cline{3-12}
    {\bfseries (\%)} &  {\bfseries (\%)}    &   {\bfseries Block}   &   {\bfseries Merges}  &   {\bfseries  Block}  &   {\bfseries Page}    &   {\bfseries Merges}  &   {\bfseries Stages}  &   {\bfseries  Block}  &   {\bfseries Page}    &   {\bfseries Merges}  &   {\bfseries Stages}  \\
    \hline
    \hline
    \multirow{8}{*}{50\%}   &   1\%     &   85690   &   209 &   2394    &   76133   &   480 &   209 &   2050    &   138731  &   5   &   209 \\
    &       &   (100\%) &   &   (3.05\%)    &       &       &       &   (1.46\%)    &       &       &       \\
    \cline{2-12}
        &   2\% &   36490   &   89  &   2120    &   55690   &   426 &   89  &   1640    &   103789  &   4   &   89  \\
    &       &   (100\%) &       &   (3.67\%)    &       &       &       &   (1.55\%)    &       &       &       \\
    \cline{2-12}
        &   5\%     &   11890   &   29  &   1665    &   40435   &   335 &   29  &   1230    &   77974   &   3   &   29  \\
    &       &   (100\%) &       &   (3.95\%)    &       &       &       &   (1.55\%)    &       &       &       \\
    \cline{2-12}
        &   10\%    &   5330    &   13  &   1569    &   33026   &   315 &   13  &   1230    &   64846   &   3   &   13  \\
    &       &   (100\%) &   &   (4.53\%)    &       &       &       &   (1.86\%)    &       &       &       \\
    \hline
    \hline
    \multirow{8}{*}{25\%}   &   1\%     &   85690   &   209 &   4083    &   65517   &   455 &   209 &   3690    &   138731  &   9   &   209 \\
    &       &   (100\%) &   &   (5.86\%)    &       &       &       &   (2.59\%)    &       &       &       \\
    \cline{2-12}
        &   2\%     &   36490   &   89  &   3492    &   51186   &   392 &   89  &   2870    &   103792  &   7   &   89  \\
    &       &   (100\%) &   &   (6.38\%)    &       &       &       &   (2.69\%)    &       &       &       \\
    \cline{2-12}
        &   5\%     &   11890   &   29  &   2871    &   38995   &   323 &   29  &   2460    &   77977   &   6   &   29  \\
    &       &   (100\%) &   &   (6.85\%)    &       &       &       &   (3.05\%)    &       &       &       \\
    \cline{2-12}
        &   10\%    &   5330    &   13  &   2419    &   32390   &   271 &   13  &   2050    &   64845   &   5   &   13  \\
    &       &   (100\%) &   &   (6.94\%)    &       &       &       &   (3.06\%)    &       &       &       \\
    \hline
    \hline
    \multirow{8}{*}{12.5\%} &   1\%     &   85690   &   209 &   7421    &   60108   &   441 &   209 &   6970    &   138733  &   17  &   209 \\
    &       &   (100\%) &   &   (10.99\%)   &       &       &       &   (4.78\%)    &       &       &       \\
    \cline{2-12}
        &   2\%     &   36490   &   89  &   6307    &   48883   &   375 &   89  &   5740    &   103796  &   14  &   89  \\
    &       &   (100\%) &   &   (11.43\%)   &       &       &       &   (5.24\%)    &       &       &       \\
    \cline{2-12}
        &   5\%     &   11890   &   29  &   5134    &   38250   &   306 &   29  &   4920    &   77982   &   12  &   29  \\
    &       &   (100\%) &   &   (11.83\%)   &       &       &       &   (5.93\%)    &       &       &       \\
    \cline{2-12}
        &   10\%    &   5330    &   13  &   4404    &   32066   &   264 &   13  &   4100    &   64848   &   10  &   13  \\
    &       &   (100\%) &   &   (12.07\%)   &       &       &       &   (5.95\%)    &       &       &       \\
    \hline
    \hline

    \end{tabular}
    \end{small}
    \caption{Block and Page level Operations on Wiki.}     
    \label{tab:writes}
\end{table*}
To better understand the I/O performance, we drill down a bit
further on some of the core operations within the I/O subsystem for
all three methods. Table \ref{tab:writes} shows the block and page
level operations, number of merges and number of stages for each
method for varying RAM buffer percentages and varying change segment
sizes on the Wikipedia dataset. For each method, "{\bf Block}"
represents the number of {\em block} operations
 and "{\bf Page}" represents the number of {\em page} level operations.
The percentage information with the Block value represents the ratio
of Block level operations to total number of (Block + Page)
operations. Columns "{\bf Merges}" and "{\bf Stages}" list the
number of merges and stages for a method, respectively.  Note that
{\em MB} has no page level operations and does not leverage staging.

Before we discuss the main trends we should highlight that the
number of merges for each of the methods is quite different. It
should be noted that a merge operation for each of the methods is
not exactly the same. Recall that a merge for {\em MB} essentially
entails unifying the contents of the memory buffer with the data
segment.  In fact the number of merge operations for {\em MB} is
identical to the number of stage operations for the other two
methods. Staging in both these methods involve unifying the contents
of the memory buffer with the change segment via sequential writes
or semi-random writes. A merge operation for {\em MDB-L} entails
unifying the entire contents of the change segment with the data
segment. A merge operation for {\em MDB} entails merging the
contents of one block within the change segment (that has filled up)
with the contents of the data segment. This explains the difference
in the number of merges for these methods.

The main trends we observe from this table are noted below.

\begin{enumerate}
\item The number of basic I/O operations (Block, Page) and meta-level
      I/O operations (Merges, Stages) (and therefore the total
      I/O cost) drops as we increase the RAM buffer
      size. The rationale for this is obvious.

    \item The ratio of block level operations to page level operations
      increases both with increasing RAM buffer size and with
      reducing the change segment size for both {\em MDB} and {\em MDB-L}.
      The number of stages decreases faster than the number of merges if the RAM buffer is increased.  If the change segment size is reduced, the number of merges will increase because the change segment will fill more frequently.

    \item {\em MDB-L} typically has the lowest number of block level operations
      while {\em MB} requires the larges number of block operations (which is
     significantly more expensive than page operations. \footnote{MLC-1 is on the order of 30 and 50 times more expensive for block level reads and block level writes, MLC-2 is over 25 and 35, and SLC is over 24 and 28 respectively.}    The low block
     operations of {\em MDB-L} can be attributed to the linear
     change segment.

     \item In terms of overall I/O costs {\em MDB} and
          {\em MDB-L} have a similar profile while
       {\em MB} is significantly more expensive.

\end{enumerate}

Summing up the I/O performance it is fair to say that for most
reasonable parameter settings {\em MDB} and {\em MDB-L}
significantly outperform {\em MB} in terms of the cost of I/O from
the perspective of the flash device. Additionally it should be noted
that the merging operation within both {\em MDB} and {\em MDB-L}
will happen completely within the SSD (allowing for an overlap of
CPU operations -- not reflected in any of the experiments) whereas a
merge for {\em MB} and staging for the other two methods will
require some CPU intervention.  Also note that an {\em MB} merge
operation is significantly more expensive than and {\em MDB} or {\em
MDB-L} stage operation (random writes versus semi-random/sequential
writes).

 \subsection{Cleans}

In our next experiment we take a closer look at the number of clean operations
required by these methods for both datasets (see Figure~\ref{fig:cleans}).  Our graphs display the variation of RAM buffer size for both datasets and the variation of change segment size for the Wiki dataset along the Wiki dataset for the X-axis.  The Y-axis is the amount of erasures.
The main trends and explanations for these trends are:
i) the number of cleans goes down with increasing RAM buffer sizes since
there are fewer stages and merges as shown in Figure \ref{fig:cleans-wiki};
ii) the number of cleans is significantly higher for {\em MB} compared to that of the other two methods because the change segment provides an extra level of buffering for {\em MDB} and {\em MDB-L} as shown in Figure \ref{fig:cleans-meme};
iii) the number of cleans increases for {\em MDB} and {\em MDB-L} as we decrease the size of the change segment because the change segment fills more often and thus there are more merges.  {\em MB} does not use the change segment so it stays a constant value.
and iv) the number of cleans for {\em MDB} and {\em MDB-L} are very similar with {\em MDB-L} being slightly better. The reduction for {\em MDB-L} is
clearly attributable to the linear change segment design.

\section{Conclusions}
\label{sec:conclusions} Hash tables pose a challenge for flash
devices.  Updating an entry inside of a disk hash table may trigger
an entire erasure of an SSD block.  Repeatedly updating a hash table
can be detrimental to the limited lifetime of the underlying SSD. A
simple hash table {\em without} buffering can be implemented.  It
has superior query time but it induces a substantial amount of
cleans and I/O cost.  From our experiments, we believe that an SSD
friendly hash table will have a RAM buffer and a disk based buffer
that supports semi-random writes. These features will increase the
locality of updates and reduce the I/O cost of the hash table for
both low and high end SSDs.  Overall our results reveal that when
one accounts for both I/O performance and query performance, {\em
MDB-L} seems to offer the best of both worlds on the workloads we
evaluated for reasonable parameter settings of change segment size
and RAM buffer size.  Furthermore, MLC-2 seems to offer the best
trade-off currently when taking into account both cost and
performance of the device. In the future, we would also like to
extend our design to hash functions that do not rely on the {\em
mod} operator (e.g. extendible hashing) and examine various
checkpointing methods.

\bibliographystyle{abbrv}
\bibliography{clemonst}

\begin{thebibliography}{10}

\bibitem{aboulnaga}
A.~Aboulnaga, A.~R. Alameldeen, and J.~F. Naughton.
\newblock {Estimating the Selectivity of XML Path Expressions for Internet
  Scale Applications}.
\newblock In {\em VLDB}, 2001.

\bibitem{agrawal}
D.~Agrawal, D.~Ganesan, R.~Sitaraman, Y.~Diao, and S.~Singh.
\newblock {Lazy-adaptive tree: An optimized index structure for flash devices}.
\newblock {\em VLDB}, 2009.

\bibitem{agrawal2}
N.~Agrawal, V.~Prabhakaran, T.~Wobber, J.~Davis, M.~Manasse, and R.~Panigrahy.
\newblock {Design tradeoffs for SSD performance}.
\newblock In {\em USENIX}, 2008.

\bibitem{cams}
A.~Anand, C.~Muthukrishnan, S.~Kappes, A.~Akella, and S.~Nath.
\newblock {Cheap and Large CAMs for High Performance Data-Intensive Networked
  Systems}.
\newblock {\em USENIX}, 2010.

\bibitem{fawn}
D.~Andersen, J.~Franklin, M.~Kaminsky, A.~Phanishayee, L.~Tan, and
  V.~Vasudevan.
\newblock {FAWN: A Fast Array of Wimpy Nodes}.
\newblock In {\em SOSP}, 2009.

\bibitem{birrell}
A.~Birrell, M.~Isard, T.~C., and T.~Wobber.
\newblock {A Design for High Performance Flash Disks}.
\newblock {\em SIGOPS}, 2007.

\bibitem{disksim}
J.~S. Bucy, J.~Schindler, S.~W. Schlosser, G.~R. Ganger, and et~al.
\newblock The disksim simulation environment version 4.0 reference manual, May
  2008.

\bibitem{cormen}
T.~H. Cormen, C.~E. Leiserson, R.~L. Rivest, and C.~Stein.
\newblock {\em {Introduction to Algorithms 2nd edition}}.
\newblock The MIT Press, 2001.

\bibitem{dai2004elf}
H.~Dai, M.~Neufeld, and R.~Han.
\newblock {ELF: an efficient log-structured flash file system for micro sensor
  nodes}.
\newblock In {\em SenSys}, 2004.

\bibitem{cstash}
B.~Debnath, S.~Sengupta, and J.~Li.
\newblock {ChunkStash: Speeding up Inline Storage Deduplication using Flash
  Memory }.
\newblock In {\em USENIX}, 2010.

\bibitem{fstore}
B.~Debnath, S.~Sengupta, and J.~Li.
\newblock {FlashStore: High Throughput Persistent Key-Value Store}.
\newblock In {\em VLDB}, 2010.

\bibitem{friedman}
S.~Friedman, P.~Krishnamurthy, R.~Chamberlain, R.~K. Cytron, and J.~E. Fritts.
\newblock {Dusty Caches for Reference Counting Garbage Collection}.
\newblock In {\em MEDEA}, 2005.

\bibitem{graefe}
G.~Graefe.
\newblock {The Five-minute Rule Twenty Years Later, and How Flash Memory
  Changes the Rules}.
\newblock {\em DaMoN}, 2007.

\bibitem{hpPaper}
L.~Hewlett-Packard Development~Company.
\newblock {Solid state drive technology for ProLiant servers: Technology
  brief}, October 2008.

\bibitem{hyun2010}
C.~Hyun, Y.~Oh, E.~Kim, J.~Choi, D.~Lee, and S.~Noh.
\newblock {A Performance Model and File System Space Allocation Scheme for
  SSD}.
\newblock {\em MSST}, 2010.

\bibitem{tfidf3}
K.~S. Jones.
\newblock A statistical interpretation of term specificity and its application
  in retrieval.
\newblock {\em Journal of Documentation}, 28:11--21, 1972.

\bibitem{kang}
D.~Kang, D.~Jung, J.-U. Kang, and K.~J-S.
\newblock {$\mu$-tree: An ordered index structure for NAND flash memory}.
\newblock {\em ICESS}, 2007.

\bibitem{kim2009flashlite}
H.~Kim and U.~Ramachandran.
\newblock {FlashLite: A User-Level Library to Enhance Durability of SSD for P2P
  File Sharing}.
\newblock In {\em ICDCS}, 2009.

\bibitem{kim2002space}
J.~Kim, J.~Kim, S.~Noh, S.~Min, and Y.~Cho.
\newblock {A space-efficient flash translation layer for compactflash systems}.
\newblock {\em ICCE}, 2002.

\bibitem{KimWS10}
Y.-R. Kim, K.-Y. Whang, and I.-Y. Song.
\newblock Page-differential logging: an efficient and dbms-independent approach
  for storing data into flash memory.
\newblock In {\em SIGMOD}, 2010.

\bibitem{koltsidas}
I.~Koltsidas and S.~D. Viglas.
\newblock {Flashing Up the Storage Layer}.
\newblock {\em VLDB}, 2008.

\bibitem{lee1}
S.~Lee and B.~Moon.
\newblock {Design of flash-based DBMS: an in-page logging approach}.
\newblock {\em SIGMOD}, 2007.

\bibitem{lee}
S.~Lee, B.~Moon, C.~Park, J.-M. Kim, and S.-W. Kim.
\newblock {A case for flash memory SSD in enterprise database applications}.
\newblock {\em SIGMOD}, 2008.

\bibitem{levanoni}
Y.~Levanoni and E.~Petrank.
\newblock {An on-the-fly reference-counting garbage collector for java}.
\newblock {\em TOPLAS}, 2006.

\bibitem{li}
Y.~Li, B.~He, Q.~Luo, and K.~He.
\newblock {Tree Indexing on Flash Disks}.
\newblock {\em ICDE}, 2009.

\bibitem{faisal1}
Y.~Li, B.~He, R.~J. Yang, Q.~Luo, and K.~Yi.
\newblock Tree indexing on solid state drives, 2010.

\bibitem{ssdAdOn}
Microsoft.
\newblock {SSD Extension for DiskSim Simulation Environment}, March 2009.

\bibitem{nath}
S.~Nath and P.~Gibbons.
\newblock {Online maintenance of very large random samples on flash storage}.
\newblock {\em VLDB}, 2010.

\bibitem{nath2}
S.~Nath and A.~Kansal.
\newblock {FlashDB: dynamic self-tuning database for NAND flash}.
\newblock {\em IPSN}, 2007.

\bibitem{relue}
R.~Relue and X.~Wu.
\newblock {Rule Generation With the Pattern Repository}.
\newblock In {\em ICAIS}, 2002.

\bibitem{rosenblum1992design}
M.~Rosenblum and J.~Ousterhout.
\newblock {The design and implementation of a log-structured file system}.
\newblock {\em TOCS}, 1992.

\bibitem{tfidf2}
G.~Salton and C.~Buckley.
\newblock Term-weighting approaches in automatic text retrieval.
\newblock In {\em INFORMATION PROCESSING AND MANAGEMENT}, pages 513--523, 1988.

\bibitem{cplus}
B.~Stroustrup.
\newblock {\em The C++ Programming Language}.
\newblock Addison-Wesley Professional, 3rd edition, June 1997.

\bibitem{woodhouse2001jffs}
D.~Woodhouse.
\newblock {JFFS: The journalling flash file system}.
\newblock Red Hat, 2001.

\bibitem{cwu}
C.-H. Wu, T.-W. Kuo, and L.-P. Chang.
\newblock {An efficient B-tree layer implementation for flash-memory storage
  systems}.
\newblock {\em TECS}, 2007.

\bibitem{mhash}
D.~Zeinalipour-Yatzi, V.~Kalogeraki, D.~Gunopulos, and W.~Najjar.
\newblock {MicroHash: An Efficient Index Structure for Flash-Based Sensor
  Devices}.
\newblock In {\em In FAST}, 2005.

\bibitem{zweig}
G.~Zweig, P.~Nguyen, J.~Droppo, and A.~Acero.
\newblock {Continuous Speech Recognition with a TF-IDF Acoustic Model}.
\newblock In {\em ISCA}, 2007.

\end{thebibliography}

\end{document}